\documentclass[cits]{PoS}
\usepackage[utf8]{inputenc}
\usepackage[usenames]{color}
\let\normalcolor\relax
\usepackage{nicefrac}
\usepackage{multirow}
\usepackage{subfig}

\usepackage{amsmath}
\usepackage{amssymb}
\usepackage{graphicx}
\usepackage{dsfont}
\usepackage{dcolumn}
\usepackage{units}
\usepackage{bm}

      \def\Slash#1{\setbox0=\hbox{$#1$} 
      \dimen0=\wd0 
      \setbox1=\hbox{/} \dimen1=\wd1 
      \ifdim\dimen0>\dimen1 
      \rlap{\hbox to \dimen0{\hfil/\hfil}} 
      #1 
      \else 
      \rlap{\hbox to \dimen1{\hfil$#1$\hfil}} 
      / 
      \fi}

      \def\Journal#1#2#3#4{{#1} {#2} (#4) #3 }

      \def\PLB{{\em Phys. Lett.} B}

      \def\PRL{\em Phys. Rev. Lett.}

      \def\PRD{{\em Phys. Rev.} D}
      \def\PRC{{\em Phys. Rev.} C}

      \def\ANNP{\em Ann. Phys.}

      \newcommand{\conjg}[1]{\ensuremath{\hspace{1pt}\overline{\hspace{-1pt}#1\hspace{-1pt}}}\hspace{1pt}}

      \def\longlonglongrightarrow{
      \relbar\joinrel\relbar\joinrel\relbar\joinrel\relbar\joinrel\relbar\joinrel\relbar\joinrel\rightarrow}
      \def\longlonglonglongrightarrow{
      \relbar\joinrel\relbar\joinrel\relbar\joinrel\relbar\joinrel\relbar\joinrel\relbar\joinrel\relbar\joinrel\relbar\joinrel\relbar\joinrel\rightarrow}

      \title{Baryon form factors \\ from Dyson-Schwinger equations}

      \ShortTitle{Baryon form factors from Dyson-Schwinger equations}

      \author{Gernot Eichmann \\
              Institut f\"ur Theoretische Physik, Justus-Liebig-Universit\"at Giessen, D-35392 Giessen, Germany\\
              E-mail: \email{gernot.eichmann@theo.physik.uni-giessen.de}}

      \abstract{ I briefly summarize the application of the Dyson-Schwinger/Faddeev approach to baryon form factors.
                 Recent results for nucleon electromagnetic and axial form factors
                 as well as $N\Delta\gamma$ electromagnetic transition form factors are discussed.
                 The calculation of the current diagrams from the quark--gluon level
                 enables an analysis of common features, such as the implications of dynamical chiral symmetry breaking and
                 quark orbital angular momentum, the timelike structure of the form factors,
                 and their interpretation in terms of missing pion-cloud effects.}

      \FullConference{International Workshop on QCD Green's Functions, Confinement and Phenomenology\\
      		 5-9 September 2011\\
      		 Trento, Italy}

\begin{document}

   \section{Introduction}

        Probing hadrons with electromagnetic, axial and pseudoscalar currents reveals their basic structure properties
        and provides a connection with the underlying quark and gluon dynamics in
        Quantum Chromodynamics (QCD). While the nucleon's axial structure is experimentally more difficult to access,
        an abundance of information has been collected for photon-induced processes
        that are described by $NN\gamma$ elastic and $N\Delta\gamma$ transition form factors.
        Precision measurements  have stimulated the development of tools to
        address questions related to quark orbital angular-momentum correlations in the perturbative domain,
        the transition between perturbative and non-perturbative regions, or pion-cloud
        rescattering effects in the chiral and low-momentum region.
        The associated chiral non-analyticities stemming from the nucleon's 'pion cloud' have been frequently
        discussed when connecting results from lattice QCD, chiral effective field theories and quark models with experiment.

        A complementary framework for studying hadron phenomenology is the one via Dyson-Schwinger equations (DSEs).
        They interrelate QCD's Green functions and provide access to nonperturbative
        phenomena such as dynamical chiral symmetry breaking and confinement, see~\cite{Alkofer:2000wg,Fischer:2006ub} for reviews.
        The investigation of hadron structure in the Dyson-Schwinger approach
        proceeds via covariant bound-state equations, i.e., the Bethe-Salpeter equation (BSE) for mesons
        and the covariant Faddeev equation for baryons~\cite{Chang:2011vu,Eichmann:2009qa}.
        The approach has several benefits:
        it is Poincar\'e-covariant throughout every step and
        provides access to all momentum scales and all quark masses without the need for extrapolations.
        Since one operates directly with QCD's degrees of freedom, observable phenomena at the hadron level
        can be systematically traced back to their microscopic origin.

        The drawback of the approach is its necessity of truncations.
        Owing to the numerical complexity of the Faddeev equation, present baryon calculations have
        been performed in a rainbow-ladder truncation, where
        $qqq$ interactions are neglected and the $qq$ and $q\bar{q}$ interactions
        are modeled by a dressed gluon exchange.
        As a consequence, several phenomenologically important features are missed in the resulting form factors.
        A characteristic example  is the absence of pion-cloud contributions in their chiral and low-momentum structure.
        The relevant gluon topologies that generate pion-cloud effects at the hadron level
        are not captured by a rainbow-ladder truncation which therefore represents the baryon's 'quark core'.
        In the case of the $N\Delta\gamma$ transition form factors discussed below, an additional quark-diquark
        simplification is made, where scalar and axialvector diquark correlations approximate the $qq$ scattering matrix
        and lead to an effective two-body description.

        In the following we will summarize recent results for the nucleon's electromagnetic,
        axial and $N\Delta\gamma$ transition form factors.
        More detailed discussions, result tables as well as references to experimental
        and lattice data which are frequently used in the plots for comparison can be found in
        Refs.~\cite{Eichmann:2011vu,Eichmann:2011pv,Eichmann:2011rw}.

   \section{Covariant Faddeev approach}

        The description of baryon structure properties in the Dyson-Schwinger approach requires knowledge of
        the nucleon and $\Delta$ bound-state wave functions and their microscopic ingredients in terms of QCD's Green functions.
        A convenient starting point is given by the three-quark connected and amputated scattering matrix $\mathbf{T}$.
        It encodes the relevant information on baryons which correspond to poles in $\mathbf{T}$.
        At a given pole for a baryon with mass $M$ the scattering matrix assumes the form
        \begin{equation}\label{poles-in-T}
           \mathbf{T}\stackrel{P^2=-M^2}{\longlonglongrightarrow} \frac{\Psi\,\conjg{\Psi}}{P^2+M^2}
        \end{equation}
        which defines the baryon's covariant wave function $\Psi$, and $\conjg{\Psi}$ is its charge conjugate\footnote{For simplicity, we will use the term
        'wave function' here both for the bound-state \textit{amplitude} $\Psi$ and the bound-state \textit{wave function} $G_0 \Psi$, where the latter has quark propagator legs attached.}.

        \begin{figure}[t]
                    \centering

                    \includegraphics[scale=1.03]{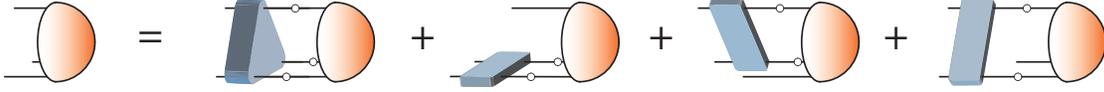}
                    \caption{Covariant three-body equation for a baryon wave function, cf.~Eq.~(2.3).}
                    \label{fig:faddeev}

        \end{figure}

        Of course, if the scattering matrix were known a priori, the masses of nucleon and $\Delta$ together with their
        wave functions could be directly extracted.
        This is not the case and thus we aim for relations that allow
        to circumvent the explicit determination of $\mathbf{T}$ in practical applications.
        $\mathbf{T}$ satisfies a scattering equation, \textit{i.e.}, the nonperturbatively resummed Dyson series
        \begin{equation}\label{dyson-eq}
             \mathbf{T} = \mathbf{K} + \mathbf{K} \,G_0 \,\mathbf{T} \,, \qquad \text{with} \quad
             G_0 = S \otimes S \otimes S \quad \text{and} \quad
             \mathbf{K} = K_\text{[3]} + \sum_{a=1}^3 S^{-1}_{(a)}\otimes K_{(a)}\,.
        \end{equation}
        By construction, the kernel $\mathbf{K}$ is the sum of a three-quark irreducible contribution
        $K_{[3]}$ and permuted two-quark irreducible kernels $K_{(a)}$,
        where the subscript $a$ stands for the respective spectator quark. $S$ denotes the dressed quark propagator.
        The combination of Eqs.~(\ref{poles-in-T}--\ref{dyson-eq}), evaluated at a bound-state pole $P^2=-M^2$,
        yields a self-consistent integral equation for the baryon wave function, cf.~Fig.~\ref{fig:faddeev}:
          \begin{equation}\label{three-quark-eq}
              \Psi = \mathbf{K}\,G_0\,\Psi\,.
          \end{equation}
        It can be solved once the dressed quark propagator and the $qq$ and $qqq$ kernels,
        which encode the interactions at the quark-gluon level, are determined.
        Naturally, all these relations are equally valid in the \textit{meson} case if the three-quark scattering matrix and kernel
        are replaced by their $q\bar{q}$ analogues and $G_0$ is taken as the $q\bar{q}$ propagator product.

        The coupling of the baryon to an external $q\bar{q}$ current, on the other hand, is reflected by the 'gauged'
        scattering matrix $\mathbf{T}^\mu$ whose residue at the bound-state pole defines the current matrix element $J^\mu$:
           \begin{equation}\label{poles-in-Tmu}
               \mathbf{T}^\mu \stackrel{P_i^2=P_f^2=-M^2}{\longlonglonglongrightarrow}   - \frac{\Psi_f\,J^\mu \,\conjg{\Psi}_i}{(P_f^2+M^2)(P_i^2+M^2)}\,.
           \end{equation}
        Depending on the type of current, which we generically denote by the index $\mu$, the respective matrix element $J^\mu$ contains
        for example the electromagnetic, axial or pseudoscalar form factors of the baryon. Here $P_i$ and $P_f$ are the incoming and outgoing
        baryon momenta and $\Psi_i$ and $\Psi_f$ are the corresponding wave functions.
        They need not describe the same type of baryon; for instance, Eq.~\eqref{poles-in-Tmu} could also be applied to the $N\Delta\gamma$ transition.

        \begin{figure}[t]
                    \centering

                    \includegraphics[scale=1.53]{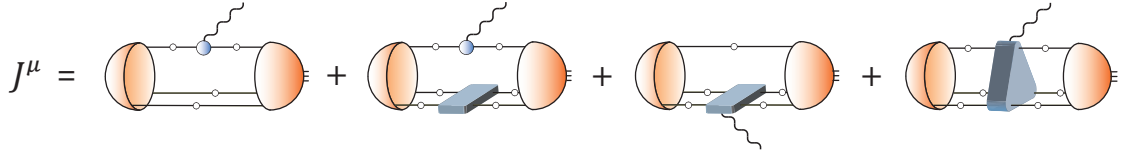}
                    \caption{General expression for a baryon's current matrix element given in Eq.~(2.5).
                    The $q\bar{q}$ vertex, dressed quark propagator, and $qq$ and $qqq$ kernels are sandwiched between incoming
                    and outgoing baryon wave functions.}
                    \label{fig:current}

        \end{figure}

        The requirement that the current couples linearly to all internal building blocks of the scattering matrix $\mathbf{T}$ implies
        that it has the formal properties of a derivative. Eq.~\eqref{dyson-eq} can then be used to resolve $\mathbf{T}^\mu$
        to a coupling to the dressed quark propagator and the kernel $\mathbf{K}$.
        The general expression for a baryon's non-perturbative current is thereby obtained as follows~\cite{Kvinikhidze:1999xp,Oettel:1999gc,Eichmann:2011vu}:
         \begin{equation}\label{emcurrent-gauging}
            J^\mu = \conjg{\Psi}_f \left(\mathbf{T}^{-1}\right)^\mu \Psi_i
                      = \conjg{\Psi}_f \,G_0 \left(\boldsymbol{\Gamma}^\mu - \mathbf{K}^\mu \right) G_0 \,\Psi_i \,.
         \end{equation}
        It is illustrated in Fig.~\ref{fig:current}
        and consists of an impulse-approximation diagram and further contributions involving the $qq$ and $qqq$ kernels.
        Its ingredients are given by
        \begin{equation}\label{current-ingredients}
               \boldsymbol{\Gamma}^\mu = \sum_{a=1}^3 \Gamma^\mu_{(a)} \otimes S^{-1}_{(b)} \otimes S^{-1}_{(c)}\,, \qquad
               \mathbf{K}^\mu = \sum_{a=1}^3 \Gamma^\mu_{(a)} \otimes K_{(a)}  + \sum_{a=1}^3 S^{-1}_{(a)} \otimes K_{(a)}^\mu + K_\text{[3]}^\mu,
        \end{equation}
        where the quark labels $\{a,b,c\}$ are an even permutation of $\{1,2,3\}$.

        Let us characterize the external current (for example, electromagnetic, axialvector or pseudoscalar) by
        $\Gamma^\mu_0 \in \left\{ \, Z_2  i\gamma^\mu, \;  Z_2 \gamma_5 \gamma^\mu , \; Z_4 i\gamma_5 \, \right\}$,
        equipped with appropriate flavor structures,
        where $Z_2$ and $Z_4$ are renormalization constants.
        The microscopic coupling of the current to the quark is then represented by the respective
        $q\bar{q}$ vertex $\Gamma^\mu$ which satisfies a Dyson-Schwinger equation:
        \begin{equation}\label{ibse}
           \Gamma^\mu = \Gamma^\mu_0 + \mathbf{T}\,G_0\,\Gamma^\mu_0  =   \Gamma^\mu_0 + K\,G_0\,\Gamma^\mu\,,
        \end{equation}
        where $\mathbf{T}$ denotes now the $q\bar{q}$ scattering matrix and $K$ the $q\bar{q}$ kernel.
        Pictorially speaking, this amounts to the sum of a pointlike part plus
        all possible reaction mechanisms between quark and antiquark which constitute the scattering matrix.
        In the second step we have exploited the scattering equation~\eqref{dyson-eq} for $\mathbf{T}$
        to obtain an inhomogeneous Bethe-Salpeter equation for the vertex which,
        in analogy to the bound-state equation~\eqref{three-quark-eq},
        allows to determine the vertex self-consistently from the $q\bar{q}$ kernel $K$.

        The appearance of the quark-antiquark $\mathbf{T}$-matrix in the defining equation for $\Gamma^\mu$ entails that the vertex
        contains meson poles whenever the bare structure $\Gamma^\mu_0$ has non-vanishing overlap with the respective
        meson wave function, cf.~Eq.~\eqref{poles-in-T}:
        \begin{equation}\label{G-Pole2}
            \Gamma^\mu \stackrel{Q^2 \rightarrow -m_\text{M}^2}{\longlonglongrightarrow} \Psi_\text{M}\,\frac{r_\text{M}^\mu}{Q^2+m_\text{M}^2}\,, \qquad
            r_\text{M}^\mu  = \text{Tr}\int \conjg{\Psi}_\text{M}\,G_0\,\Gamma_0^\mu \,\Big|_{Q^2\rightarrow -m_\text{M}^2}\,.
        \end{equation}
        Here, $Q=P_f-P_i$ is the total $q\bar{q}$ momentum that flows into the vertex and the index 'M' stands for meson.
         Since $\Gamma^\mu$ enters the form factors diagrams via Eq.~\eqref{current-ingredients}
        it is clear that these poles must also appear in the timelike $Q^2$ structure of the form factors where they
        set the relevant scales: the $\rho-$meson dominates electromagnetic processes, the axialvector meson $a_1$
        appears in axial (isovector) form factors, and the pion and its excitations in pseudoscalar form factors of hadrons.

        The same reasoning was generalized in Ref.~\cite{Eichmann:2011ec} to derive
        the hadron's coupling to \textit{two} external currents with $q\bar{q}$ quantum numbers:
        \begin{equation}
        \begin{split}
            & \quad \mathbf{T}^{\mu\nu} \stackrel{P_i^2=P_f^2=-M^2}{\longlonglonglongrightarrow}  \frac{\Psi_f\,J^{\mu\nu} \,\conjg{\Psi}_i}{(P_f^2+M^2)(P_i^2+M^2)}\,, \\
            & J^{\mu\nu} = \conjg{\Psi}_f \left[ \left(\mathbf{T}^{-1}\right)^{\{\mu} \mathbf{T} \left(\mathbf{T}^{-1}\right)^{\nu\}} - \left(\mathbf{T}^{-1}\right)^{\mu\nu} \right] \Psi_i \,,
        \end{split}
        \end{equation}
        where the curly brackets denote symmetrization of the indices.
        Depending on the types of hadrons and currents involved, the resulting scattering amplitudes $J^{\mu\nu}$ can describe a variety
        of different reactions such as Compton scattering, pion electroproduction, $N\pi$ or $\pi\pi$ scattering, or crossed-channel
        processes such as $p\bar{p}$ annihilation into two photons or meson production.

        We have now outlined a systematic approach to compute various hadron properties from their underlying non-perturbative substructure in QCD.
        Its input is provided by the quark propagator $S$ and the $qq$, $q\bar{q}$ and $qqq$ kernels. Once these quantities are determined,
        no further model input is required: we can selfconsistently solve
        the bound-state equation~\eqref{three-quark-eq} to obtain a hadron's wave function and mass,
        solve the inhomogeneous BSE~\eqref{ibse} for the vertex, and combine them to calculate hadron form factors and scattering amplitudes.
        The Dirac-Lorentz structure of the wave functions, vertices and current matrix elements is fully determined from Poincar\'e covariance.
        By implementing the complete 'operator basis' in each case, its momentum-dependent Lorentz-invariant dressing functions (the 'form factors')
        are obtained from the equations described above.

        In order to proceed, we have to specify a truncation procedure.
        The apparent problem is the lack of information on the kernels which,
        according to the reasoning so far, encode the information from QCD's Green functions
        that is relevant for hadron physics.
        Omitting the term $K_{[3]}$ in Eqs.~\eqref{dyson-eq} and~\eqref{current-ingredients}
        yields the covariant \textit{Faddeev equation} together with its corresponding currents~\cite{Eichmann:2009qa,Eichmann:2011vu}.
        They trace the binding mechanism of three quarks in a baryon to its quark-quark correlations.
        Moreover, the simplest ansatz for the $qq$ (and $q\bar{q}$) kernel is the rainbow-ladder kernel:
             \begin{equation}\label{RLkernel}
                 K =  Z_2^2 \, \frac{ 4\pi \alpha(k^2)}{k^2} \, T^{\mu\nu}_k \gamma^\mu \otimes \gamma^\nu,
             \end{equation}
        where $T^{\mu\nu}_k=\delta^{\mu\nu} - k^\mu k^\nu/k^2$ is a transverse projector with respect to the gluon momentum $k$.
        Its implementation in the Faddeev equation yields, by iteration, all dressed-gluon ladder exchanges between quark pairs.
        Implemented in the current, only the direct couplings to the quarks,\textit{ i.e.}, the first two terms in Fig.~\ref{fig:current}, survive.

        The rainbow-ladder kernel satisfies vector and axialvector Ward-Takahashi identities
        which ensure electromagnetic current conservation and the Gell-Mann-Oakes-Renner
        and Goldberger-Treiman relations at the hadron level~\cite{Maris:1997hd,Eichmann:2011pv}.
        Through these identities, Eq.~\eqref{RLkernel} also determines the kernel of the quark DSE
        whose solution is numerically straightforward.
        The quark-gluon vertex is thereby reduced to its vector structure $\sim\gamma^\mu$, and its nonperturbative dressing,
        together with that of the gluon propagator, is absorbed in an effective interaction $\alpha(k^2)$ which is the remaining model input.
        We employ the Maris-Tandy ansatz of Ref.~\cite{Maris:1999nt} which reproduces the one-loop logarithmic running at large gluon momenta
        and features a Gaussian bump in the infrared that generates dynamical chiral symmetry breaking.
        Different parametrizations have been recently tested and yield similar results
        for a range of hadron properties~\cite{Blank:2010pa,Qin:2011dd,SanchisAlepuz:2011jy}.

        \begin{figure}[t]
                    \centering

                    \includegraphics[scale=1.77]{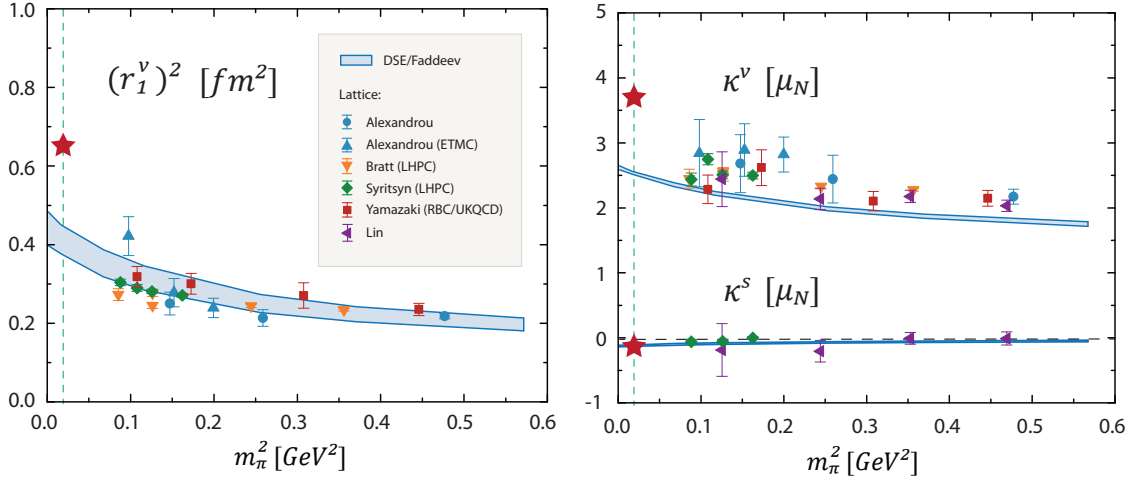}
                    \caption{Quark-mass dependence of nucleon static electromagnetic properties compared to lattice results.
                             \textit{Left panel:} squared isovector Dirac radius $(r_1^v)^2$.
                             \textit{Right panel:} isovector and isoscalar anomalous magnetic moments $\kappa^v$ and $\kappa^s$ in units of nuclear magnetons.
                             Stars denote the experimental values.
                             Figure adapted from Ref.~\cite{Eichmann:2011vu}.}\label{fig:nucleon-emffs-static}

        \end{figure}

  \section{Nucleon electromagnetic form factors \label{sec:ffs-nucleon-em}}

       The nucleon's electromagnetic current is expressed by two dimensionless form factors:
       the Dirac and Pauli form factors $F_1(Q^2)$ and $F_2(Q^2)$, or the Sachs form factors $G_E(Q^2)$ and $G_M(Q^2)$ as their linear combinations:
       $G_E = F_1 - Q^2/(4M_N^2)\, F_2$ and $G_M = F_1 + F_2$. The current matrix element is given by
       \begin{equation}
           J^\mu = i\Lambda^+_f \left[ F_1(Q^2) \,\gamma^\mu - F_2(Q^2)\,\frac{\sigma^{\mu\nu} Q^\nu}{2M_N}  \right] \Lambda^+_i\,,
       \end{equation}
       where $Q=P_f-P_i$ is the photon momentum and $\Lambda^+_{i,f} = \tfrac{1}{2}\,\big(\mathds{1} + \widehat{\Slash{P}}_{i,f}\big)$ are positive-energy projectors.
       In the static limit one retrieves the nucleons' anomalous magnetic moments $\kappa = F_2(0)$
       as well as their Dirac and Pauli radii $r_1^2 = -6F_1'(0)$ and $r_2^2 = -6F_2'(0)/F_2(0)$.
       The isoscalar (isovector) form factors are the sum (difference) of proton and neutron form factors: $F_i^{s,v} = F_i^p \pm F_i^n$.

\begin{figure}[t]
                    \centering

                    \includegraphics[scale=1.02]{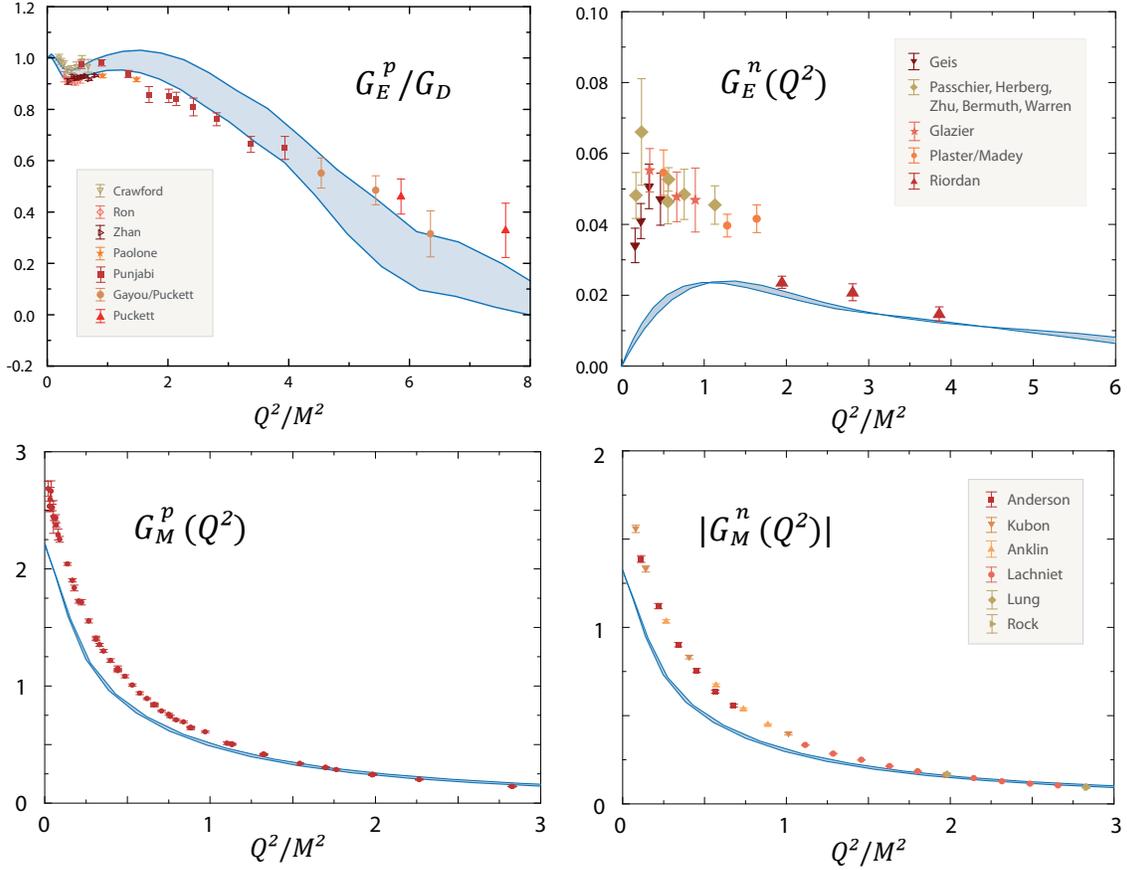}
                    \caption{Nucleon electromagnetic Sachs form factors as functions of the photon momentum transfer and in comparison with experimental data.
                             The proton's electric form factor in the top left panel is normalized by the standard dipole.
                             Figure adapted from Ref.~\cite{Eichmann:2011vu}. }\label{fig:nucleon-emffs}

        \end{figure}

       Results for the pion-mass dependence and $Q^2-$evolution of various nucleon electromagnetic form factors
       are shown in Figs.~\ref{fig:nucleon-emffs-static} and~\ref{fig:nucleon-emffs}.
       The bands correspond to a variation of the infrared properties in the quark-gluon interaction $\alpha(k^2)$ and measure the model uncertainty.
       As anticipated, the absence of pion-cloud contributions in the chiral and low-momentum region is recovered in the results.
       All form factors are in reasonable agreement with experimental data at larger momentum transfer
       where the nucleon is probed at small length scales and the pion cloud becomes irrelevant.
       Missing structure mainly appears in the low-momentum region $Q^2 \lesssim 2$~GeV$^2$.
       The calculated charge radii, such as the isovector Dirac radius in the left panel of Fig.~\ref{fig:nucleon-emffs-static},
       underestimate their experimental values but converge with lattice data at larger quark masses.
       Pion loops would increase the charge radii toward the chiral limit where they would diverge.

       Chiral effective field theory predicts that leading-order chiral corrections to proton and neutron
       anomalous magnetic moments carry an opposite sign;
       their magnitude is therefore enhanced in the isovector combination $\kappa^v = \kappa^p-\kappa^n$
       and cancels in the isoscalar case $\kappa^s = \kappa^p+\kappa^n$.
       The isoscalar magnetic moment is quite accurately reproduced by the Faddeev calculation: $\kappa^s = -0.12(1)$,
       compared to the experimental value $\kappa^s_\text{exp} = -0.12$~\cite{Eichmann:2011vu}.
       The calculated values of $\kappa^s$ and $\kappa^v$ correspond to an underestimation
       of $20\%-30\%$ in the proton and neutron magnetic moments $G^{p,n}_M(0)$,
       visible in the bottom panels of Fig.~\ref{fig:nucleon-emffs}.
       Another example is the neutron electric form factor $G_E^n(Q^2)$ in Fig.~\ref{fig:nucleon-emffs} which
       agrees with recent measurements at larger $Q^2$ but misses the characteristic bump at low $Q^2$.
       These observations suggest to identify the rainbow-ladder truncated nucleon with the 'quark core' in chiral effective field theories.

       The large--$Q^2$ behavior of form factors is of great theoretical and experimental interest as well.
       The experimental falloff of the proton's form factor ratio $G_E^p/G_M^p$ has been
       attributed to orbital angular-momentum correlations in the nucleon wave function
       which modify the perturbative scaling behavior and entail a zero crossing in $G_E^p(Q^2)$.
       Quark orbital angular momentum in terms of $s$, $p$ and $d$ waves appears in the Dirac-Lorentz structure of the nucleon's rest-frame Faddeev amplitude.
       While nucleon and $\Delta$ baryons are dominated by $s$ waves, $p$ waves play an important role as well:
       they contribute $\sim 30\%$ to the nucleon's canonical normalization and diminish only slowly with increasing current-quark masses.
       The contribution from $d$ waves, on the other hand, is below $1\%$.
       At large $Q^2$, the form-factor results from the Faddeev calculation become sensitive to the numerics; nevertheless,
       a decrease of $G_E^p$ compared to the dipole form is visible in Fig.~\ref{fig:nucleon-emffs} and implies a zero crossing as well.

       Another remark concerns the timelike behavior of the form factors and the
       vector-meson dominance property which is a consequence of the underlying dynamics.
       The electromagnetic current is microscopically represented by the quark-photon vertex which can be separated
       in two terms: a Ball-Chiu part that satisfies electromagnetic gauge invariance,
       and another purely transverse term that includes vector-meson poles in the $J^{PC}=1^{--}$ channel~\cite{Ball:1980ay,Maris:1999bh}.
       Since the rainbow-ladder truncation does not dynamically develop hadronic decay widths,
       the poles that are generated in the self-consistent calculation of the quark-photon vertex are timelike and real.
       The decomposition into 'Ball-Chiu' and '$\rho$-meson' contributions can be made in all electromagnetic hadron form factors which
       therefore possess poles at $Q^2=-m_\rho^2$ and further $1^{--}$ excited-state locations.
       The transverse term is negative at spacelike $Q^2$ and, in the case of electric form factors, vanishes at $Q^2=0$, i.e.,
       the Ball-Chiu part alone satisfies charge conservation $G_E^p(0)=1$. The $\rho-$meson term
       contributes roughly $\sim 50\%$ to the nucleon's squared charge radii
       throughout the current-mass range but has only a minor impact on its magnetic moments
       whose overall contribution comes from the Ball-Chiu term.

       We note that a reduction of the Faddeev equation to a quark-diquark description, where scalar and axialvector diquark correlations are
       calculated from the same quark-gluon input, yields quite similar results for the form factors~\cite{Eichmann:2009zx}.
       The model dependence is however larger, especially at large $Q^2$,
       and the corresponding bands in Fig.~\ref{fig:nucleon-emffs} become sizeable; cf. also Fig.~\ref{fig:REM-RSM} below.
       Nevertheless, these results imply that the interaction of quarks with
       scalar and axialvector diquarks provides  the overwhelming contribution to the nucleon's binding.

       \section{Nucleon axial form factors \label{subsec:ffs-ax}}

       In Ref.~\cite{Eichmann:2011pv}, the Faddeev approach
       has been further applied to compute the nucleon's axial and pseudoscalar form factors.
       The respective current matrix elements are specified by the axial form factor $G_A(Q^2)$,
       the induced pseudoscalar form factor $G_P(Q^2)$, and the pseudoscalar form factor $G_5(Q^2)$:
       \begin{equation}
           J^\mu_5 =  \Lambda^+_f \gamma_5 \left[ G_A(Q^2) \,\gamma^\mu + G_P(Q^2)\, \frac{i \,Q^\mu}{2M_N} \right] \Lambda^+_i\,, \qquad
           J_5 = G_5(Q^2)\,\Lambda^+_f\,i\gamma_5\,\Lambda^+_i.
       \end{equation}
       Their microscopic decomposition in the Faddeev framework is identical to Fig.~\ref{fig:current} except for the type of $q\bar{q}$ vertices that are involved:
       the structure $\gamma^\mu$ that enters the self-consistent calculation of the quark-photon vertex
       is replaced by $\gamma_5 \gamma^\mu $ and $\gamma_5$, respectively.
       Again, the pole structure of the resulting axial and pseudoscalar vertices
       allows to extract information on the timelike behavior and identify the relevant scales in the form factors.
       $G_A$ is dominated by the $1^{++}$ axialvector meson $a_1(1260)$ and its excitations whereas
       $G_P$ and $G_5$ are governed by the pion pole. The pion-nucleon form factor $G_{\pi NN}$ is the residue of $G_5$ at the pion pole
       and thus related to the $\pi(1300)$ and further $0^{-+}$ excitations.
       The Goldberger-Treiman relation $G_A(0) = f_\pi\,G_{\pi NN}(0)/M_N$
       follows as a consequence of the axialvector Ward-Takahashi identity and analyticity which are satisfied microscopically.

\begin{figure}[t]
                    \begin{center}

                    \includegraphics[scale=1.55]{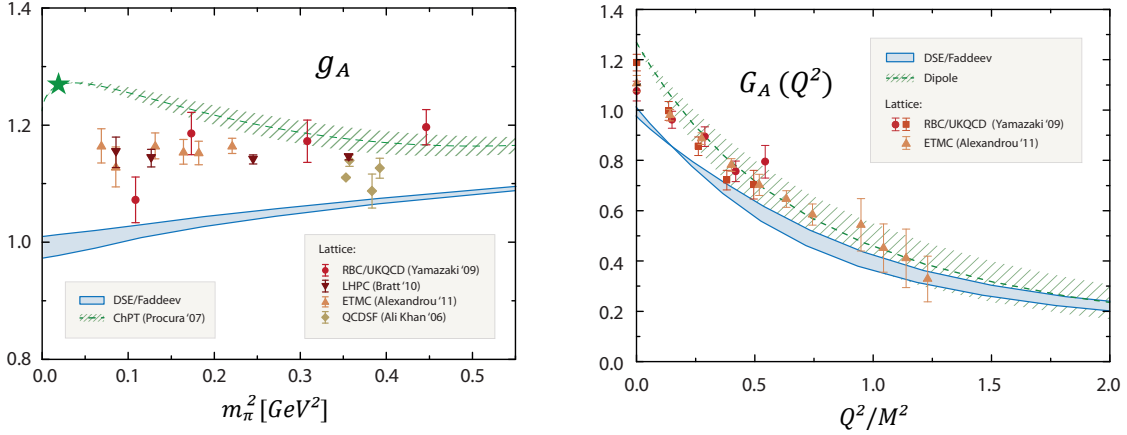}
                    \caption{\textit{Left panel:} Quark-mass dependence of the nucleon's axial charge $g_A$, compared to lattice results and the chiral expansion of Ref.~\cite{Procura:2006gq}.
                             \textit{Right panel:} $Q^2-$evolution of the axial form factor $G_A(Q^2)$, compared to lattice data and the experimental dipole form.
                             Figure adapted from Ref.~\cite{Eichmann:2011pv}.}\label{fig:GA}

                    \end{center}
        \end{figure}

       The (isovector) axial and pseudoscalar form factor results exhibit various similarities with their electromagnetic counterparts, see Fig.~\ref{fig:GA}.
       The axial charge $g_A = G_A(0)$ underestimates the experimental value by $20\%-25\%$; it falls below recent lattice data in
       the low quark-mass region and approaches the chiral expansion at larger pion masses.
       On the other hand, $G_A(Q^2)$ is consistent with the phenomenological dipole form at larger $Q^2$.
       Analogous results are obtained for the remaining pseudoscalar form factors.
       This suggests once again that these features are signals of missing pion-cloud effects.
       Such an interpretation was also proposed to explain the volume dependence of lattice results for $g_A$~\cite{Ohta:2011nv}.

       \section{Electromagnetic $N\to\Delta$ transition \label{sec:ndeltagamma}}

       Finally, the approach can be applied for the calculation of $\Delta(1232)$ and $N\to\Delta$ transition
       form factors as well. Since a solution for the $\Delta$ bound-state amplitude
       from the Faddeev equation has become available only recently~\cite{SanchisAlepuz:2011jn},
       we will restrict our discussion to the quark-diquark model.
       The derivation that leads to the diagrams in Fig.~\ref{fig:current} yields
       analogous expressions in the quark-diquark approach~\cite{Oettel:1998bk,Oettel:1999gc},
       where the diquark ingredients can be computed self-consistently from the same quark-gluon input.
       Form-factor results in that framework exist for nucleon and $\Delta$ electromagnetic form factors~\cite{Eichmann:2009zx,Nicmorus:2010sd} and
       the $\Delta N\pi$ pseudoscalar transition~\cite{Mader:2011zf}, and in the following we will summarize
       recent results for the electromagnetic $N\Delta\gamma$ transition~\cite{Eichmann:2011rw}.

        \begin{figure}[t]
                    \centering
                    \includegraphics[scale=1.65]{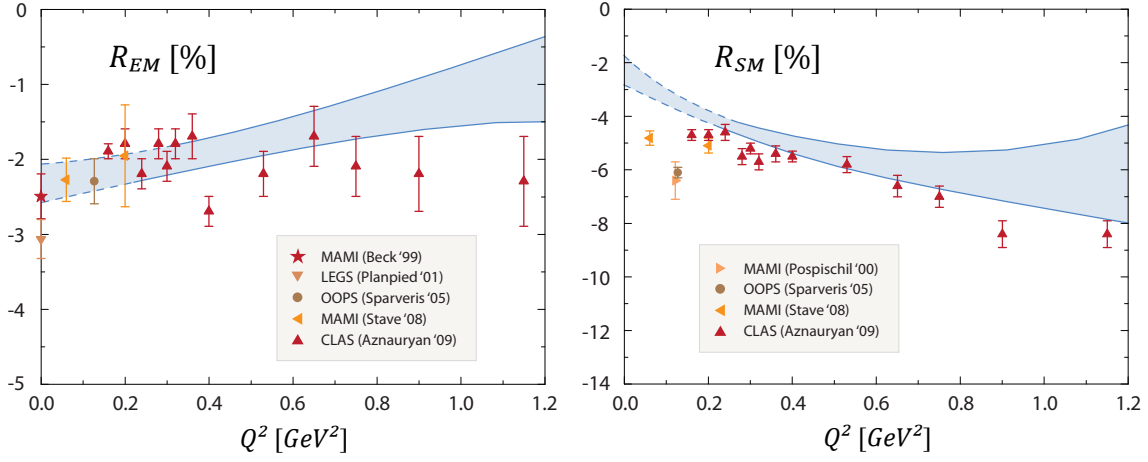}
                    \caption{$Q^2-$dependence of the electric and Coulomb quadrupole form-factor ratios $R_{EM}$ and $R_{SM}$ compared to experimental data.
                             Figure adapted from Ref.~\cite{Eichmann:2011rw}.}\label{fig:REM-RSM} \vspace{5mm}
        \end{figure}

       The $N\Delta\gamma$ transition is characterized by the three Jones-Scadron form factors
       $G_M^\star(Q^2)$, $G_E^\star(Q^2)$ and $G_C^\star(Q^2)$ which are related to the
       pion electroproduction multipole amplitudes at the $\Delta-$resonance position~\cite{Jones:1972ky,Pascalutsa:2006up}.    
       The respective current $J^{\mu,\rho}$ is decomposed as follows:
       \begin{equation}\label{current-general-0b}
           J^{\mu,\rho} = b\,\mathds{P}^{\rho\alpha}_f \,i\gamma_5
           \left[ \frac{i\omega}{2\lambda_+}\,(G_M^\star-G_E^\star)\,\gamma_5 \,\varepsilon^{\alpha\mu\gamma\delta} K^\gamma \widehat{Q}^\delta
            - G_E^\star \, T^{\alpha\gamma}_Q \,T^{\gamma\mu}_K - \frac{i\tau}{\omega}\,G_C^\star\,\widehat{Q}^\alpha K^\mu \right] \Lambda^+_i\,.
       \end{equation}
       Instead of the incoming and outgoing momenta $P_i$ and $P_f$, we used the orthonormal four-momenta $\widehat{Q}^\mu$ and $K^\mu = \widehat{P_T}^\mu$, where
       a hat denotes normalization, $P  = (P_f+P_i)/2$ is the average momentum and $P_T$ its component transverse to the photon momentum.
       The Rarita-Schwinger projector for the $\Delta-$baryon reads
       \begin{equation}
           \mathds{P}^{\rho\alpha}_f = \Lambda^+_f \,T^{\rho\sigma}_{P_f} \left( \delta^{\sigma\beta}  - \gamma^\sigma \gamma^\beta\right) T^{\beta\alpha}_{P_f}\,,
       \end{equation}
       and the transverse projectors $T^{\mu\nu}_{P_f}$, $T^{\mu\nu}_Q$ and $T^{\mu\nu}_K$ are defined in the same way
       as in the gluon case below Eq.~\eqref{RLkernel}.
       The remaining dimensionless variables in~\eqref{current-general-0b} are given by:
       \begin{equation}\label{tau-lambda}
           \tau := \frac{Q^2}{2\,(M_\Delta^2+M_N^2)}  \,, \quad
           \lambda_\pm := \frac{(M_\Delta \pm M_N)^2 + Q^2}{2\,(M_\Delta^2+M_N^2)}\,, \quad
           \omega:= \sqrt{\lambda_+ \lambda_-}\,, \quad
           b :=\sqrt{\frac{3}{2}} \left(1 + \frac{M_\Delta}{M_N}\right).
       \end{equation}

       The $N\Delta\gamma$ transition has been accurately measured over a wide momentum range~\cite{Pascalutsa:2006up,Aznauryan:2011qj}.
       It is dominated by a magnetic dipole transition which, in a quark-model picture, amounts to the spinflip of a quark
       and is related to the form factor $G_M^\star(Q^2)$.
       The remaining electric and Coulomb quadrupole form factors are much smaller and expressed by
       the ratios $R_{EM}(Q^2)$ and $R_{SM}(Q^2)$ which encode the deformation in the transition.
       In non-relativistic quark models, non-zero values for these ratios would require $d-$wave components in the nucleon and $\Delta$ wave functions.
       On the other hand, the analysis of pion electroproduction data via dynamical reaction models suggests that $R_{EM}$ and $R_{SM}$ are almost
       entirely dominated by the pion cloud~\cite{JuliaDiaz:2006xt}.

        \begin{figure}[t]
                    \includegraphics[scale=1.14]{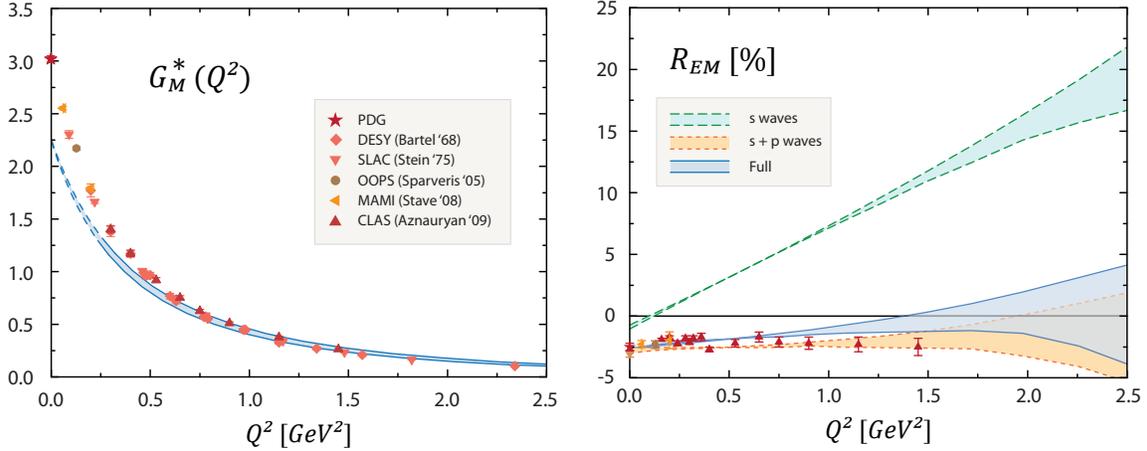}
                    \caption{\textit{Left panel:} $Q^2-$dependence of the magnetic dipole transition form factor $G_M^\star(Q^2)$
                             compared to experiment. \emph{Right panel:} decomposition of $R_{EM}(Q^2)$ according to the
                             orbital angular-momentum content in the nucleon and $\Delta$ wave functions.
                             Figure adapted from Ref.~\cite{Eichmann:2011rw}.}\label{fig:GM-REM}
        \end{figure}

       In contrast, the quark-diquark results which are plotted in Fig.~\ref{fig:REM-RSM}
       reproduce the experimental data for $R_{EM}$ and $R_{SM}$ quite well, even without the inclusion of pion-cloud corrections.
       In the case of $R_{EM}$, this behavior originates from $p-$wave contributions in the nucleon and $\Delta$ wave functions which are a consequence of
       Poincar\'e covariance. The removal of $p$ waves results in a ratio that is overall positive and grows with increasing $Q^2$, cf.~Fig.~\ref{fig:GM-REM},
       with a trend towards the perturbative prediction $R_{EM} \to 1$ for $Q^2 \to\infty$~\cite{Aznauryan:2011qj}. The impact of $d$ waves is almost negligible.

       On the other hand, the result for the magnetic dipole transition form factor $G_M^\star(Q^2)$ in Fig.~\ref{fig:GM-REM} follows the
       characteristics of the previously discussed magnetic and axial form factors: it agrees with experimental data at larger $Q^2$ and underestimates them by $\sim 25\%$
       at $Q^2 =0$. This is consistent with the quark-model result and the expected behavior of the pion cloud from coupled-channel analyses.
       Moreover, neither $G_M^\star$ nor $R_{SM}$ are sensitive to the addition of $p$ and $d$ waves but dominated by $s-$wave elements alone.

       \section{Conclusions and outlook}

       We have discussed several recent nucleon and $\Delta$ form factor results in the Dyson-Schwinger approach,
       obtained either directly from the covariant Faddeev equation or in a quark-diquark simplification.
       All calculations share the same quark-gluon input and the results display consistent features.
       Quark-quark correlations, which are mediated by a rainbow-ladder gluon-exchange interaction, can account for the overall properties
       of the nucleon and $\Delta$ quark core and justify a quark-diquark picture for these baryons.
       Dynamical chiral symmetry breaking and Poincar\'e covariance have important consequences for the behavior of the form factors.
       Their timelike structure is dominated by meson poles in the underlying quark-antiquark vertices.
       The admixture of quark orbital angular momentum via $p$ waves, even in $s-$wave dominated ground states such as the nucleon and $\Delta-$baryon, is crucial
       for the $N\Delta\gamma$ electric quadrupole form factor and the large--$Q^2$ behavior of electromagnetic form factors.
       The main missing ingredients in a rainbow-ladder approach are pion-cloud contributions at low momenta and small pion masses.

       The combination of Dyson-Schwinger and covariant bound-state equations provides valuable tools for investigating the internal structure of hadrons.
       Its applications are still at an early stage, and it is desirable to extend the framework to study more sophisticated systems and reactions such as
       baryon excitations and nucleon-to-resonance transition form factors, virtual Compton scattering, pion electroproduction,
       pion-nucleon scattering, or timelike $p\bar{p}$ annihilation processes.
       At the same time, these efforts must be complemented by technical improvements, such as residue calculus to provide kinematic access
       to truly large $Q^2$, or the implementation of pion-cloud corrections and hadronic decay channels
       via truncations beyond rainbow-ladder.

   \subsection*{\it Acknowledgements}

       I would like to thank C.~S.~Fischer and D.~Nicmorus for useful discussions.
       This work was supported by the Austrian Science Fund FWF under
       Erwin-Schr\"odinger-Stipendium No.~J3039, and the Helmholtz International Center for FAIR
       within the LOEWE program of the State of Hesse.

\end{document}